# Photon Counting and Super Homodyne Detection of Weak QPSK Signals for Quantum Key Distribution

**Q. XU[1], M.B. Costa e Silva[1], S. Agnolini[1], P. Gallion[1], F.J. Mendieta[2]**
[1] Ecole Nationale Supérieure des Télécommunications (GET/Télécom Paris, CNRS), 46 rue Barrault, 75013 Paris, France.
[2] CICESE, km.107 Carr. Tijuana, Ensenada, Baja California 22800, México
**email:** qxu@enst.fr

**Summary**
We compare the principles and experimental results of two different QPSK signal detection configurations, photon counting and super homodyning, for applications in fiber-optic Quantum Key Distribution (QKD) systems operating at telecom wavelength, using the BB84 protocol.

**Introduction**

In QKD systems where modulated optical signals are transmitted at quantum levels [1], time-multiplexed QPSK encoding [2,3] of Alice's Qbits constitutes an attractive modulation format since it reduces the requirements in phase and polarization stability. In fact differential configurations relax substantially the stability requirement in the mean square phase fluctuations of Alice's source and the required stability time is only of the order of several bits.

Several experiments have been reported that use a photon counting configuration with cooled avalanche photodiodes (APD). However these receivers have limitations due to their inherent low quantum efficiency (<0.1) and their high dark count rate which requires an operation in the gated mode, resulting in a very low key generation rate even at low operation temperature. The super homodyne detection employing conventional InGaAs P.I.N photodiodes which present a much higher quantum efficiency (>0.8) and quick response time, constitutes an interesting alternative to photon counting when a stronger reference is used to provide a high mixing gain to overcome the receiver's thermal noise. The bright coherent pulses can be used in the precise phase tracking at Bob's measurement (typically a delayed interferometer), necessary even in differential configurations.

**Discussion**

In the photon counting setup (Fig. 1), Alice and Bob apply phase modulations both in the longer arm of the two unbalanced interferometers so that the pulses of $\Phi_A$ and those of $\Phi_B$ have the same intensity at the Bob's coupler's input. Detector 1 clicks for $\Phi_A - \Phi_B = 0$ while Detector 2 clicks for $\Phi_A - \Phi_B = \pi$. When $\Phi_A - \Phi_B = \pi/2$ or $-\pi/2$, the photon arrives at Detector 1 or Detector 2 in a random way.

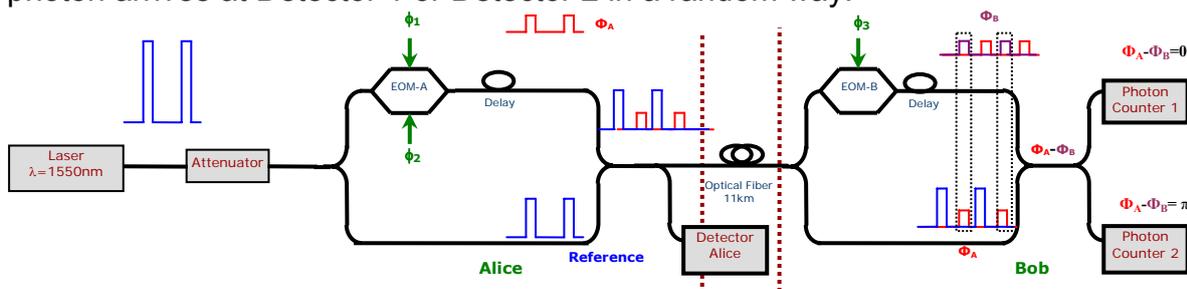

Fig. 1 QKD DQPSK Setup: Photon Counting Detection (EOM: Electro-Optical Modulator)

In our experiment the repetition frequency is set to 4MHz; the gate width is set to 2.5ns so as to minimize the dark count rate. When using a fiber interferometer without feedback from the detected signal, the mean false counts were issued mainly from: a) the unsatisfactory extinction ratio of the reference pulses which degrades the interferometer visibility; b) the polarization imperfections of the laser pulse which affect the mixing of the key pulse with the reference pulse; c) the slight deviation of modulating signals from ideal phase modulation; d) the unavoidable thermo-mechanical variations; and e) the dark counts of the APD detectors.

In the super homodyne setup (Fig. 2), we replace the photon counters by a balanced photoreceiver. At Alice's end strong reference pulses are interleaved as to have a high mixing gain when beating with the fainted pulses at Bob's end.

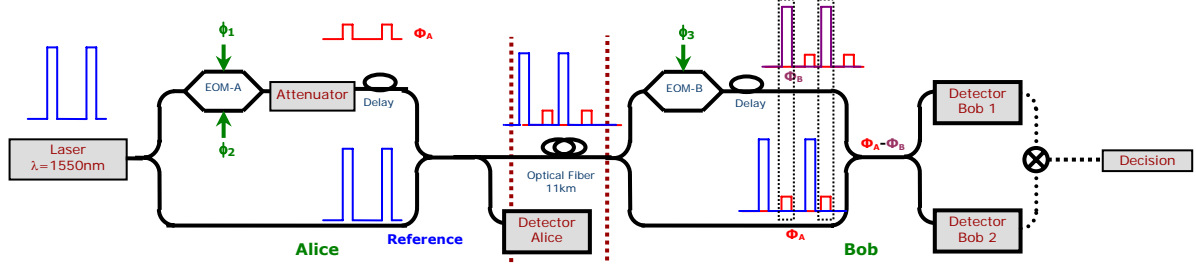

Fig. 2 QKD DQPSK Setup: Coherent Balanced Super-Homodyne Detection

In this balanced super homodyne configuration, the "shot noise" is the only limit since the strong reference pulse makes the thermal noise irrelevant ($noise_{thermal}$ = -174dBm/Hz, $noise_{shot}$ = -154dBm/Hz as measured in our experiments). In Fig. 3 we demonstrate Bob's detected symbols with 2 photons/bit. Positive pulses stand for bits "1" while negative ones stand for bits "0". Pulses of base Anti-Coincidence are discarded. When 1 photon/bit is recorded, the BER is theoretically less than 3%. Balanced configuration helps also remove common mode noises.

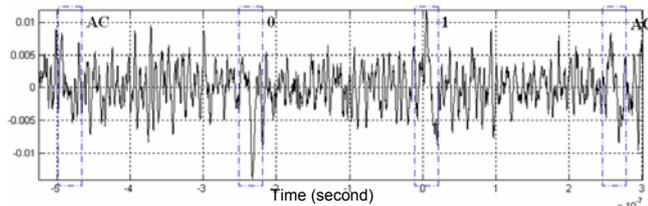

Fig. 3 Bob's detected signals (2 photons/bit)

**Conclusions**

An experimental QKD system is implemented using optical phase modulation for base and symbol encoding. While photon counting cannot sustain high bit rate due to the quenching process, super homodyne could be a good alternative to photon counting [3]. Preliminary results have been obtained although with low contrast and we are improving our experiment at the optical and electronic levels as well as on the carrier phase synchronisation and polarisation compensation.